# FiND: Few-shot three-dimensional image-free confocal focusing on point-like emitters

Swetapadma Sahoo[1,2,3], Junyue Jiang[4], Jaden Li[1,5], Kieran Loehr[3,5], Chad E. Germany[3,5], Jincheng Zhou[4], Bryan K. Clark[3,5], Simeon I. Bogdanov[1,2,3]

[1]*Department of Electrical and Computer Engineering, University of Illinois at Urbana-Champaign, Urbana, Illinois 60801, USA*

[2]*Nick Holonyak, Jr. Micro and Nanotechnology Laboratory, University of Illinois at Urbana-Champaign, Urbana, Illinois 61801, USA*

[3]*Illinois Quantum Information Science and Technology Center, University of Illinois Urbana-Champaign, Urbana, Illinois 61801, USA*

[4]*Zhejiang University-University of Illinois at Urbana-Champaign Institute, Haining 314400, China*

[5]*Department of Physics, University of Illinois Urbana-Champaign, Urbana, IL 61801, USA*

Confocal fluorescence microscopy is widely applied for the study of point-like emitters such as biomolecules, material defects, and quantum light sources. Confocal techniques offer increased optical resolution, dramatic fluorescence background rejection and sub-nanometer localization, useful in super-resolution imaging of fluorescent biomarkers, single-molecule tracking, or the characterization of quantum emitters. However, rapid, noise-robust automated 3D focusing on point-like emitters has been missing for confocal microscopes. Here, we introduce FiND (Focusing in Noisy Domain), an imaging-free, non-trained 3D focusing framework that requires no hardware add-ons or modifications. FiND achieves focusing for signal-to-noise ratios down to 1, with a few-shot operation for signal-to-noise ratios above 5. FiND enables unsupervised, large-scale focusing on a heterogeneous set of quantum emitters. Additionally, we demonstrate the potential of FiND for real-time 3D tracking by following the drift trajectory of a single NV center indefinitely with a positional precision of < 10 nm. Our results show that FiND is a useful focusing framework for the scalable analysis of point-like emitters in biology, material science, and quantum optics.

**Introduction:**

Confocal microscopes are ubiquitously used to probe fluorescent point-like emitters (PLEs) in applications such as super-resolution localization microscopy[1,2], single-particle tracking[3], ratiometric fluorescence[4], characterization of quantum optical sources [5] and defect detection in semiconductors[6]. The confocal approach offers several advantages, including reduced phototoxicity to live cells, superior signal-to-noise ratio, high-resolution imaging[7], and enables time-resolved spectroscopy[3].

3D focusing, i.e. co-locating the microscope focal point with the PLE, to obtain a near-maximum detected fluorescence intensity in real time, is critical to analyzing the properties of individual PLEs or PLE-tagged biomolecules. For example, localization accuracy and precision[8–11], as well as characterization throughput for fluorescence lifetime, or photon correlation[12], critically depend on the quality of focus. Specifically, automatic rapid focusing with sub-100 nm precision is essential for high numerical aperture (NA) confocal measurements of point-like emitters on an extensive spatial or temporal scale. Focusing in confocal systems typically involves a combination of z-focusing to place the object in the focal plane[13–15], and x-y imaging via raster-scanning[16,17]. These conventional methods are often slow, increasing phototoxicity, and the risk of photobleaching. Moreover, the time lag can introduce errors in the calculated focal position, owing to the movement of PLEs in the sample, caused by drift or diffusion[18]. Additionally, they may require extra accessory optics, such as secondary lasers[13,19,20] and detectors[15,21], thereby introducing complexity to the integration and alignment of microscope systems.

Here, we introduce FiND (Focusing in Noisy Domain)- a rapid, noise-robust framework for real-time 3D autofocusing on PLEs in confocal fluorescence microscopy. FiND is directly compatible with standard confocal microscopes, without the need for training, imaging, or additional hardware add-ons/modifications. We consider the excitation beam's focal point as a particle moving under the influence of the information and noise received from local intensity measurements. The main idea of the framework is to represent the effect of ground-truth information in the intensity measurements as an attractive force (towards the PLE) and the noise in the measurements as a repulsive force (away from the PLE) along a single dimension. We analytically predict measurement parameters that guarantee a near-unity focusing success probability and minimize the focusing time $t_{focus}$. We verify these predictions using Monte-Carlo simulations and experiments on fluorescent emitters using a laser scanning confocal fluorescence microscope. Finally, we show the applications of FiND for large-scale quantum emitter characterization and active drift correction via emitter intensity tracking.

**Results:**

We model the signal, s(**r**,t), captured by photon detector as the sum of a symmetric Gaussian[22] point spread function (PSF) g(**r**) = $e^{-\frac{r^2}{2}}$ (Fig 1(a)) and a temporally varying Gaussian noise component n(**r**,t) with standard deviation = 1/SNR. The goal of focusing is to move the objective focus into the target zone, i.e., the sphere of radius $r_\epsilon = \sqrt{2\epsilon}$ around the PLE (shaded with green in Fig. 1(b)). This zone corresponds to an average collected intensity above 1-ϵ, where ϵ represents a predefined tolerance level. For the rest of the study, we fix ϵ = 0.1, leading to $r_\epsilon$ = 0.44. A focusing attempt is considered successful if collected intensities remain above 1-ϵ, on average after initially surpassing this threshold. We hypothesize that an iterative focusing method will yield the fastest performance as its speed is fundamentally limited by the duration of a single iteration. The focusing time $t_{focus}$ is defined as the iteration number upon which the objective's position first enters the target zone.

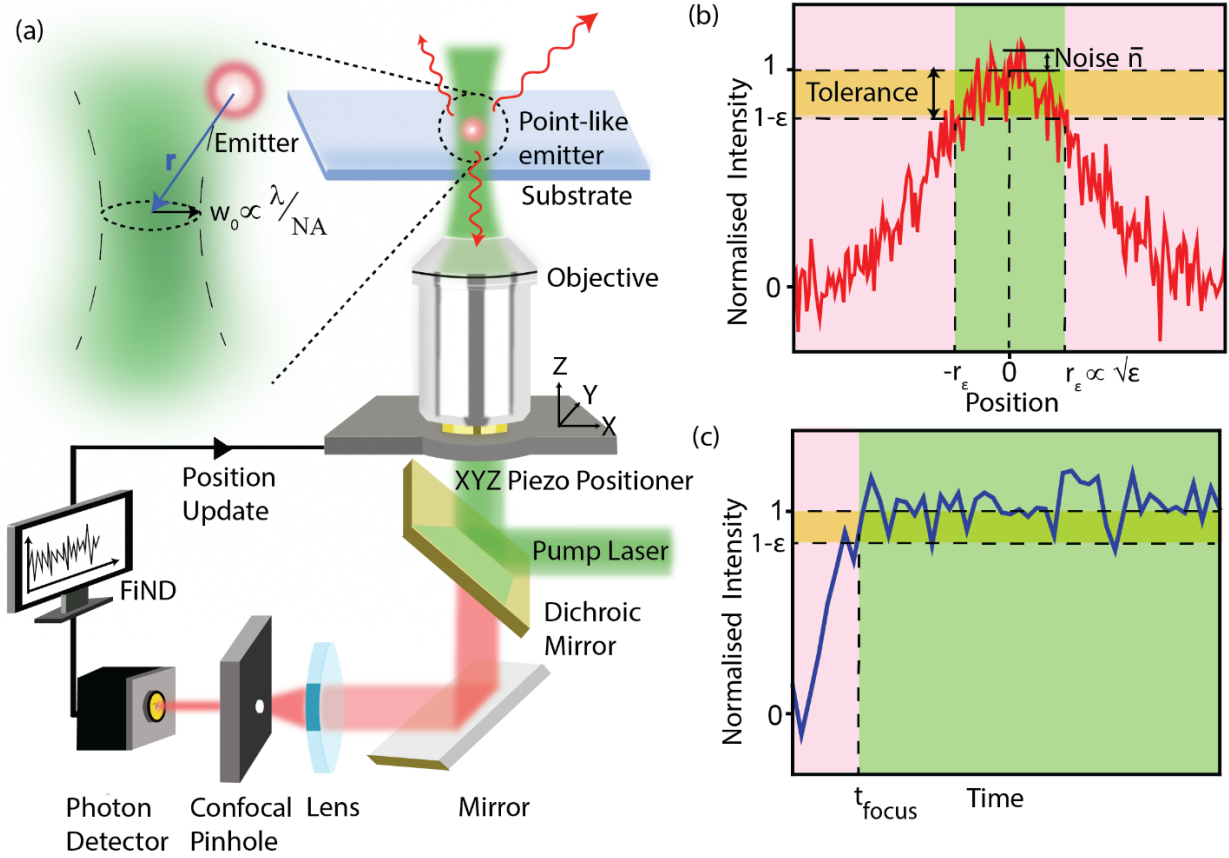

**Fig. 1.** Rapid focusing of a confocal microscope on PLEs, with a starting focus position **r** a few diffraction lengths away from the PLE (a). The system performs successive intensity measurements on a noisy Gaussian PSF, aiming to exceed a fluorescence intensity threshold of 1-ϵ. This condition corresponds to the focus position entering the target zone (green) of radius $r_\epsilon = \sqrt{2\epsilon}$ around the PLE (b). The focusing time $t_{focus}$ is the number of iterations it takes to enter the target zone (c).

In what follows, we set a framework for iterative focusing algorithms on the example of a simple finite difference method. However, the following analysis applies to any sampling strategy. Specifically, we assume for this study that for each iteration, the signal is sampled in six locations around the current position: $\mathbf{r} \pm \delta * \mathbf{e}_{x,y,z}$, where δ is referred to as the step size, and $\mathbf{e}_{x,y,z}$ are unit vectors. After sampling, the focus is displaced by the quantity $\mathbf{D} = \frac{\lambda}{2\delta}\sum_{j=x,y,z}\{s(\mathbf{r}+\delta*\mathbf{e}_j) - s(\mathbf{r}-\delta*\mathbf{e}_j)\}\mathbf{e}_j$, where λ is called the learning rate.

To analyze the performance of this scheme, we follow the focus movement along the effective radial coordinate $r^2$. We decompose the displacement **D** into the ground truth component $\mathbf{D_{GT}} = \frac{\lambda}{2\delta}\sum_{j=x,y,z}\{g(\mathbf{r}+\delta*\mathbf{e}_j) - g(\mathbf{r}-\delta*\mathbf{e}_j)\}\mathbf{e}_j$, and the noise component $\mathbf{D_N} = \frac{\lambda}{2\delta}\sum_{j=x,y,z}\{n(\mathbf{r}+\delta*\mathbf{e}_j,t) - n(\mathbf{r}-\delta*\mathbf{e}_j,t)\}\mathbf{e}_j = \mathbf{D_{TN}} + \mathbf{D_{RN}}$, where $\mathbf{D_{TN}}$ and $\mathbf{D_{RN}}$ are the transverse and radial noise components. The three displacement components add up vectorially in three dimensions. However, the contributions of these components to the effective radial coordinate $r^2$ add algebraically, which allows us to examine them intuitively as forces acting on a particle in one dimension. $\mathbf{D_{GT}}$ results in an attractive force $F_{GT}$, while $\mathbf{D_{RN}}$ and $\mathbf{D_{TN}}$ form a repulsive force $F_N$. The average resultant force $F_{res} = \langle(\mathbf{r}+\mathbf{D})^2 - \mathbf{r}^2\rangle$ can be written as (Supplementary Note 1):

$$F_{res} = F_{GT} + F_N = -2r\frac{\lambda}{\delta}e^{-\frac{r^2}{2}-\frac{\delta^2}{2}}\sinh(r\delta) + \frac{\lambda^2}{\delta^2}e^{-r^2-\delta^2}\sinh^2(r\delta) + \frac{3\lambda^2\bar{n}^2}{2\delta^2} \quad \text{(Equation 1)}$$

This expression is validated by a good quantitative agreement with Monte-Carlo simulations for SNR=1, 20 and r= $r_\epsilon$ =0.44 as a function of λ and δ (Fig 2). Slight quantitative discrepancies can be attributed to the simplifying assumptions of the analytical model, i.e., the strictly radial direction of the ground truth displacement and neglecting the higher moments of the noise displacements (Supplementary Note 1).

Eq. 1 allows to analyze the focusing performance and set optimal parameters. The focusing time can be approximated as $t_{focus} \approx \int_{r_0}^{r_\epsilon} \frac{2rdr}{F_{res}(r)}$. Numerically, we find that for r < 1.7, -$F_{res}$ increases with r, so, we choose to maximize the resultant force at the target zone boundary -$F_{res}(r_\epsilon)$, where it is expected to become the limiting factor. For $\epsilon$ = 0.1, this force is maximized by $\delta_{opt}$ = 1.03 and $\lambda_{opt} \sim \frac{0.64}{0.4+8\bar{n}^2}$ (Supplementary Note 2). Numerical simulations further show that with this simple six-point sampling scheme, FiND already outperforms natural evolution strategy (NES), particle swarm optimization (PSO), and convolutional neural network (CNN)-based curve fitting in terms of focusing speed and noise resilience (Supplementary Note 3).

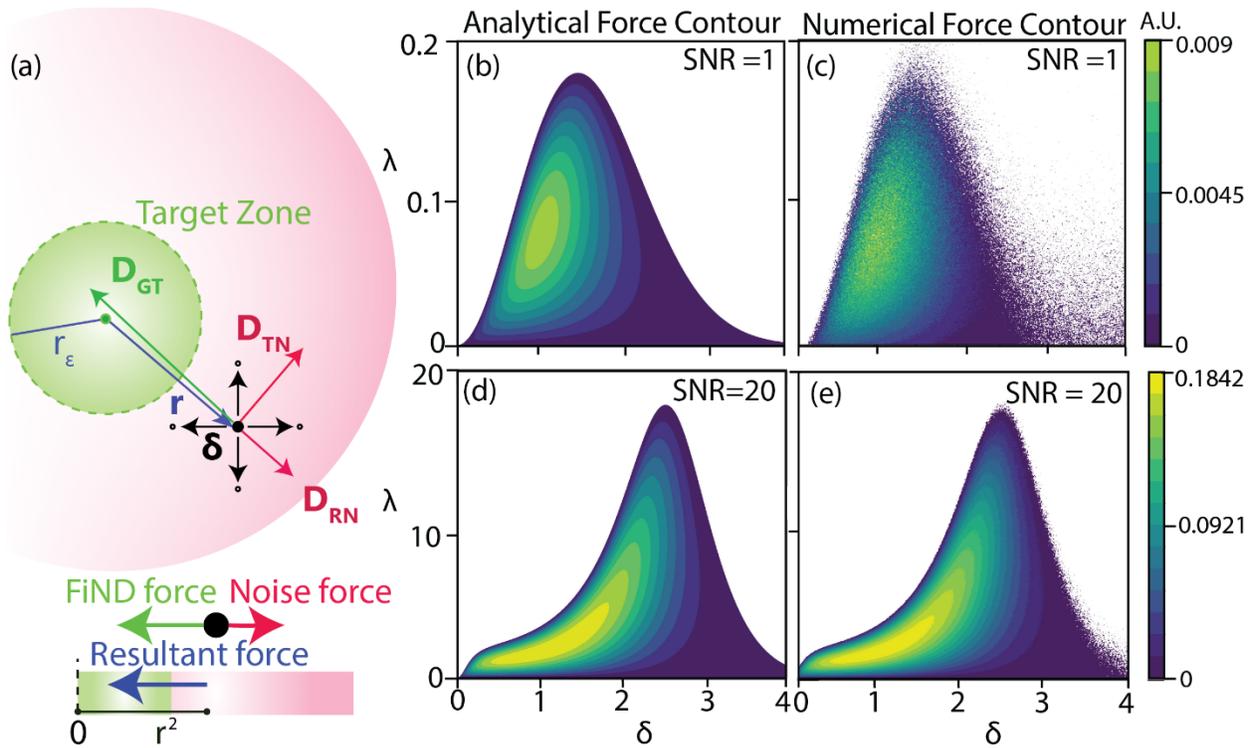

**Fig.2**. Schematic representation of the displacements in real space and forces in the 1-dimensional space. (a). Search parameters must be optimized to enter the target zone in the minimum number of iterations. Resultant forces (-$F_{res}$), as a function of λ and δ at the target zone boundary (r = $r_\epsilon$) for SNR = 1 (b,c) and SNR = 20 (d, e) calculated analytically (b,d) and numerically (c,e). White regions correspond to a repulsive resultant force.

We now experimentally benchmark the performance of the FiND framework by repeatedly focusing a standard confocal microscope on a subwavelength-size nanodiamond containing fluorescing NV centers (NV-ND) and recording $t_{focus}$ as a function of SNR and r. In line with the analysis above, the experimental radial coordinate **r_exp**, fluorescence intensity *I*, $\lambda_{exp}$ and $\delta_{exp}$ are non-dimensional. We normalize the experimental radial coordinate **r_exp**, and $\delta_{exp}$ using the gaussian PSF width $\sigma_y$ (Fig 3(a)). The collected intensity is normalized to the emitter's average intensity collected at focus: $I_{max}$. Consequently, the experimental learning rate is normalized by the quantity $\frac{\sigma^2}{I_{max}}$.

We first verify that the sign of $F_{res}(r_\epsilon)$ correctly predicts the success of focusing. We choose four different pairs ($\lambda$, $\delta$) shown on the analytical force map calculated for SNR = 26 (the experimentally retrieved value for our ND-NV) and $r = r_\epsilon$ (Fig.3(c)). In agreement with the theory, only the operating point featuring -$F_{res}(r_\epsilon)$ > 0 yields successful focusing (Fig.3(d)). Notably, the worst focusing performance is recorded for attempt #3 symbolized by the green point and corresponding to the regime commonly used in gradient search algorithms. In this regime, the noise force dominates due to the vanishing step size (Equation 1).

We now measure $t_{focus}$ as a function of SNR (Fig. 3(e)) and the starting distance from focus $r_0$ (Fig. 3(f)), for $\delta = \delta_{opt}$ and $\lambda = \lambda_{opt}$. Starting radii **r** corresponds to different starting intensities being collected by the objective, while the SNR of the ND-NV is changed by varying the laser excitation. We compare the experimental values of $t_{focus}$ (green) to those predicted by the analytical theory (black) and Monte-Carlo simulations (red) (Methods). We find a good result agreement for SNRs > 5 and r < 2, where focusing succeeds within just a few iterations.

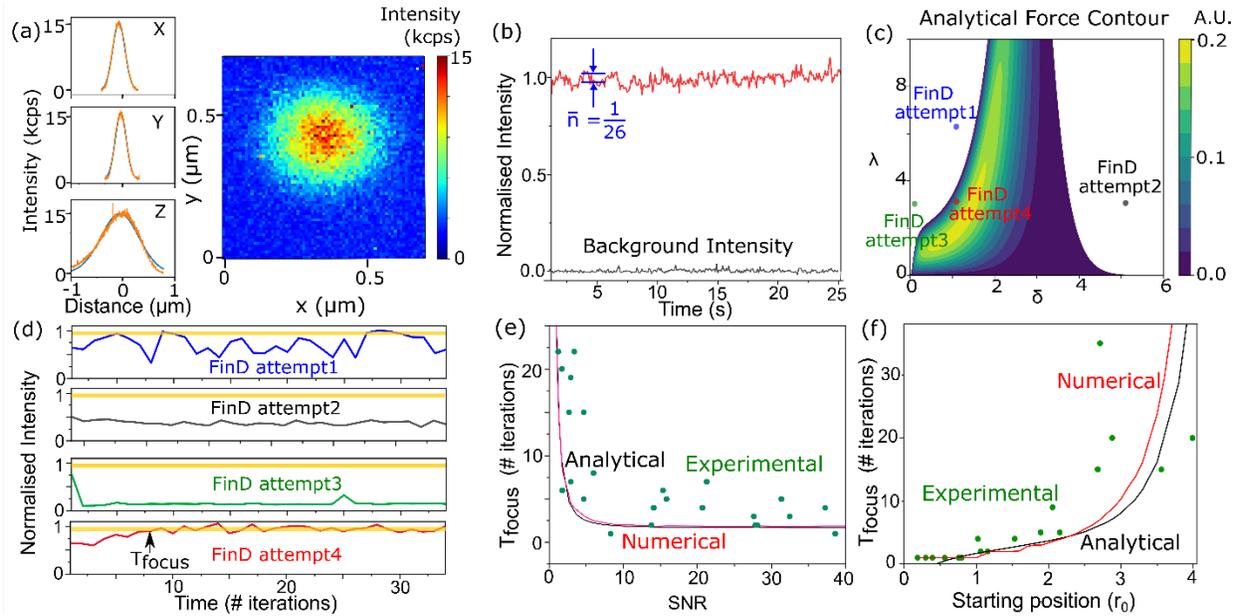

**Fig. 3.** Benchmarking of the FiND focusing framework. A photostable ND-NV emitter is chosen to investigate how the focusing time scales with forces, SNR and starting position (a) 1D PSFs taken on the manually focused emitter (excitation laser power of 300 uW) and the respective 2D PSF are shown. PSF fitting yields $\sigma_x = 0.12\ \mu m, \sigma_y = 0.1\ \mu m$ and $\sigma_z = 0.37\ \mu m$ (b) Intensity time trace of in-focus fluorescence intensity shows SNR=26 (c) The force contour plot at the target zone boundary for SNR = 26 (d) The focusing curves for the 4 parameter points chosen in (c), with the only success corresponding to a positive resultant force. Scaling of focusing time with SNR (e) and starting position (f).

Optimal focusing parameters lead to rapid focusing for a variety of SNRs and starting positions. However, in applications such as large-scale quantum emitter characterization[23] and fluorescent defect detection[24], one may need to focus successively on a set of heterogeneous and previously uncharacterized emitters. FiND allows us to tune the algorithm so that it succeeds for a wide range of emitter intensities and SNRs, without modifying λ and δ.

By analyzing the statistics of NV-ND photophysical parameters, we find that their maximal intensities and signal-to-noise ratios roughly lie in the intervals $I_{max}$ ~ 25 - 250 kilocounts per second (kcps), and SNR = 5 - 50 respectively, at an excitation laser power of 0.3 mW. We use FiND to focus on a set of a hundred NV-NDs, evaluating $t_{focus}$ for each attempt consisting of 60 iterations. The experimental step size is fixed to 0.1 and the learning rate is calculated conservatively according to the low end of the SNR range (10) and the high end of the $I_{max}$ range (200k cps). This gives a theoretical learning rate of 1.48, and an experimental learning rate of 7.44E-08 ( for σ = 0.1 μm) (Supplementary Note 2).

We observe that 90% of the emitters are focused within 26 iterations, and 97% - within 40 iterations (Fig. 4). For 3 NV-NDs with $I_{max}$ < 18 kcps, the resultant force is so weak that they fail to satisfy the success criterion within ~60 iterations.

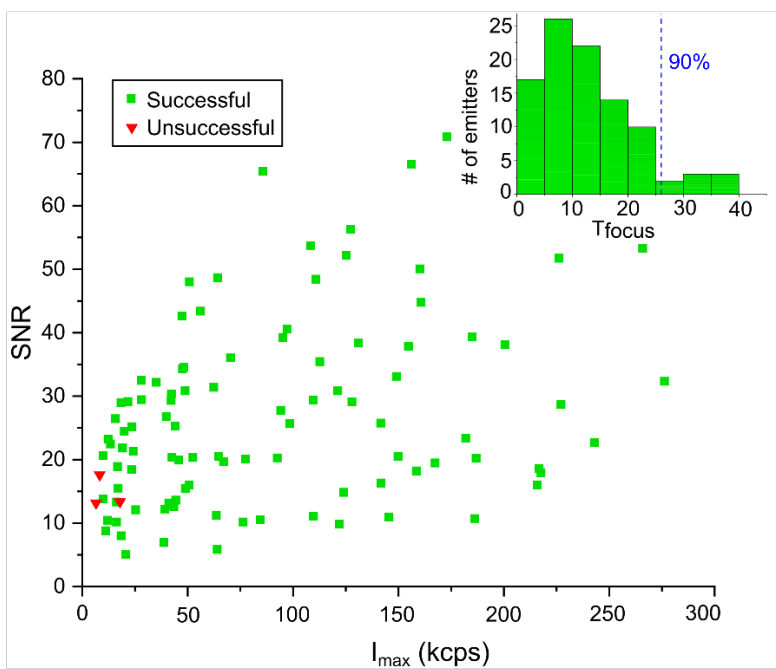

**Fig.4.** FiND-enabled large scale rapid focusing on a heterogenous set of NV-NDs with a wide distribution of maximal intensities and SNRs. Scatter plot showing successful focusing (green) at different SNRs and $I_{max}$ of ND-NV, for pre-determined parameters of FiND. Inset shows the statistics of number of emitters vs. number of iterations taken to focus. We observe that 90% of the emitters are in focus within 26 iterations (blue dashed line).

The steep intensity profile of confocal illumination can be harnessed for tracking the 3D positions of individual nanoparticles over large spatial and temporal ranges, enabling e.g. studies of cellular transport processes [3,25,26]. We evaluate the potential of FiND for real-time single-particle tracking by monitoring a photostable NV-ND's drift pattern (SNR=25), resulting from microscope drift. The red curve in Fig. 5a shows a control experiment, in which a manually focused NV-ND drifts out of focus and remains unfocused. We let FiND maintain focus using $\delta_{opt}$ and $\lambda_{opt}$ continuously. After an initial focusing period of $t_{focus}$ = 19 iterations, the average normalized photoluminescence intensity stays above 0.9 indefinitely (Fig. 5(a), green curve). In Fig. 5b we plot the piezo coordinate traces during the experiment. The uncertainties on the PLE location, estimated from nearly constant segments of the piezo coordinate traces are $\Delta_x$ ~ 9 nm, $\Delta_y$ ~ 8 nm, and $\Delta_z$ ~ 9 nm (Fig. 5(b)). These values are limited by the coordinate read-out noise of the objective's piezo nanopositioning stage and can be improved by integrating the coordinate output.

Active FiND focusing can be immediately useful for drift correction and microscope stabilization, to mitigate the effects of air currents, vibrations, and temperature variations. These factors frequently result in data distortions and reduced resolution in super-resolution techniques such as single-molecule localization microscopy[27,28] and single-particle tracking[29].

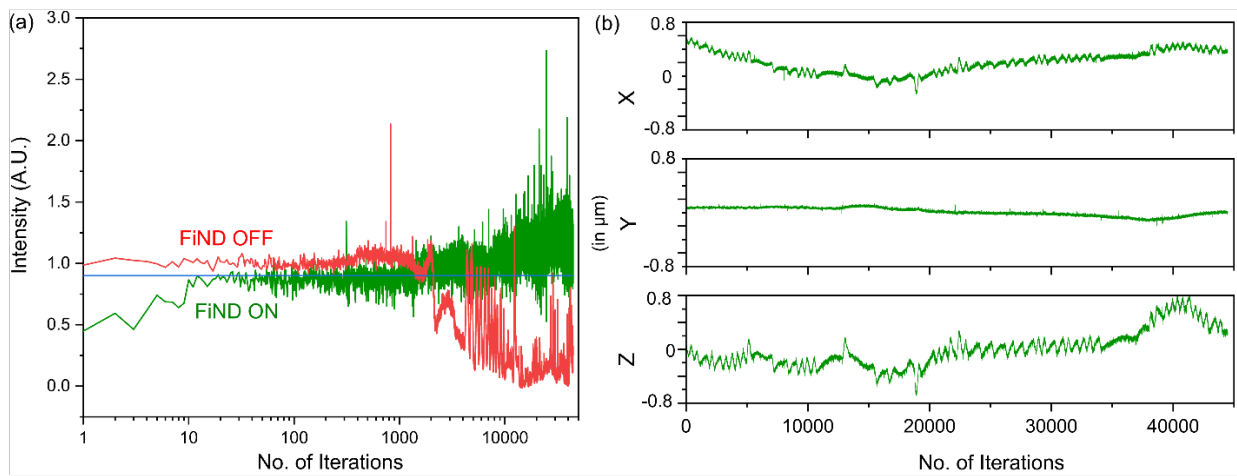

**Fig.5.** FiND maintains stable focus indefinitely by compensating for the drift on emitters. (a) When FiND is turned off, the intensity drops as the emitter drifts out of focus. The green curve shows the intensity with FiND turned on. (b) Traces of piezo coordinates when FiND is on, tracking the 3D drift with $\Delta_x$ ~ 9 nm, $\Delta_y$ ~ 8 nm and $\Delta_z$ ~ 9 nm.

### Discussion:

We developed an automated, imaging-free 3D focusing framework for maximizing intensity and active-3D tracking of point-like emitters in the confocal mode. The framework predicts optimal sampling parameters through a simple analytical model and can be applied for large-scale characterization of point-like emitters[30]. FiND tracks the real-time 3D-position of PLEs with sub-10 nm accuracy, on par or better than other diffraction-limited confocal-based intensity tracking methods[26,31,32], without requiring any additional hardware components. FiND's noise resilience allows us to operate with reduced pump powers, minimizing the risk of photobleaching in organic fluorescent biomarkers and limiting the phototoxicity in biological samples.

The memoryless nature of the iterative sampling method used here allows one to focus even on blinking emitters (Supplementary Note 4). This feature may be useful in microscopy techniques that leverage fluorophore blinking for super-resolution [33]. The performance can be further improved by exploring sample schemes beyond the 6-point finite difference and taking the asymmetric nature of the PSF into account. To improve the range of tracking speed one may use galvanoscanning mirrors, electro-optic deflectors, and acousto-optic deflectors for laser beam control[34–36].

The noise force features a moderate, linear scaling with the number of dimensions, which promises applications much beyond confocal microscopy. Gradient search optimization, widely used in machine learning, control theory, and economics, is traditionally associated with step sizes and learning rates much below unity[37]. Here, optimization of a simple single-maximum cost function in the presence of noise shows that the use of small step sizes causes sub-optimal ground truth information sampling. The use of unconventionally large learning rates and step sizes in finite difference sampling schemes can accelerate convergence rates and therefore shows potential for optimizing noisy convex-concave functions.

**Methods:**

$I_{max}$ and SNR values for each NV-ND are obtained through manual focusing and subsequently acquired 30s intensity traces. The ground truth emitter location is estimated by averaging locations from 10 FiND iterations after the normalized intensity exceeds 0.9. These locations are expected to be randomly distributed around the PLE location within the target zone. We then use this ground truth location to deduce the boundary of the target zone and calculate $t_{focus}$.

**Sample Preparation:** The sample was prepared by dropcasting a diluted aqueous suspension of 20 nm, milled fluorescent NV-NDs (Adamas Nanotechnology, NDNV20nmHi10ml) on a cleaned coverslip.

**Optical Measurements:** Optical measurements are taken using a custom-built scanning confocal microscope with a 50 $\mu$m pinhole, based on a commercial inverted microscope body (Nikon Ti-U). Focusing and scanning are done using an XYZ piezoelectric stage (PIMars Nanopositioning Stage P-561.3CD) carrying the objective. A 520 nm (green) CW laser excitation beam (OBIS 520nm LX 40mW, Coherent) was reflected off a 550 nm long-pass dichroic mirror (DMLP550L, Thorlabs) onto the back of a 100x oil objective (1.6 NA 100x/1.49 Oil objective Nikon Apo TIRF). A 550 nm long-pass filter (FEL0550, Thorlabs) was used to filter out the remaining pump power. Photoluminescence emission is collected on single-photon avalanche detectors (PDM, Micro-Photon Devices).

**Scaling law:**

To obtain the data in Fig. 3(e), the focus was moved in a random direction from a manually focused position until 60% of $I_{max}$ was being collected. The sampling parameters were $\delta = 1$ and $\lambda = \lambda_{opt} \sim \sqrt{e} \frac{2r^2}{2r^2 + (\frac{1}{SNR})^2 3e}$.

To obtain the data in Fig. 3(f), the focus was moved in a random direction from a manually focused position. Starting distance r is inferred as the distance between the coordinates of the first iteration and the ground truth emitter location (see above). The sampling parameters were $\delta = 1$ and $\lambda = 1.6$.

**Data Availability:**

The data that support the findings of this study are available from the corresponding author upon reasonable request.

**References:**

1. Masullo, L. A. *et al.* An alternative to MINFLUX that enables nanometer resolution in a confocal microscope. *Light Sci. Appl.* **11**, 199 (2022).

2. Gwosch, K. C. *et al.* MINFLUX nanoscopy delivers 3D multicolor nanometer resolution in cells. *Nat. Methods* **17**, 217–224 (2020).

3. Han, J. J., Kiss, C., Bradbury, A. R. M. & Werner, J. H. Time-Resolved, Confocal Single-Molecule Tracking of Individual Organic Dyes and Fluorescent Proteins in Three Dimensions. *ACS Nano* **6**, 8922–8932 (2012).

4. Deniz, A. A. *et al.* Ratiometric single-molecule studies of freely diffusing biomolecules. *Annu. Rev. Phys. Chem.* **52**, 233–253 (2001).

5. Breitweiser, S. A. *et al.* Efficient Optical Quantification of Heterogeneous Emitter Ensembles. *ACS Photonics* **7**, 288–295 (2020).

6. Rideout, D. Infrared Laser Confocal Microscopy: Fast, Flexible, Cost-Effective Inspection and Metrology Tool for Microelectronic Manufacturing. *Microsc. Today* **15**, 36–37 (2007).

7. Elliott, A. D. Confocal Microscopy: Principles and Modern Practices. *Curr. Protoc. Cytom.* **92**, e68 (2020).

8. Thiele, J. C. *et al.* Confocal Fluorescence-Lifetime Single-Molecule Localization Microscopy. *ACS Nano* **14**, 14190–14200 (2020).

9. Balzarotti, F. *et al.* Nanometer resolution imaging and tracking of fluorescent molecules with minimal photon fluxes. *Science* **355**, 606–612 (2017).

10. Engelhardt, J. *et al.* Molecular Orientation Affects Localization Accuracy in Superresolution Far-Field Fluorescence Microscopy. *Nano Lett.* **11**, 209–213 (2011).

11. Stallinga, S. & Rieger, B. Accuracy of the Gaussian Point Spread Function model in 2D localization microscopy. *Opt. Express* **18**, 24461–24476 (2010).

12. Kudyshev, Z. A. *et al.* Rapid Classification of Quantum Sources Enabled by Machine Learning. *Adv. Quantum Technol.* **3**, 2000067 (2020).

13. Liron, Y., Paran, Y., Zatorsky, N. G., Geiger, B. & Kam, Z. Laser autofocusing system for high-resolution cell biological imaging. *J. Microsc.* **221**, 145–151 (2006).

14. Zhang, X., Zeng, F., Li, Y. & Qiao, Y. Improvement in focusing accuracy of DNA sequencing microscope with multi-position laser differential confocal autofocus method. *Opt. Express* **26**, 887–896 (2018).

15. Jan, C.-M., Liu, C.-S. & Yang, J.-Y. Implementation and Optimization of a Dual-confocal Autofocusing System. *Sensors* **20**, 3479 (2020).

16. Shulevitz, H. J. *et al.* Template-Assisted Self-Assembly of Fluorescent Nanodiamonds for Scalable Quantum Technologies. *ACS Nano* **16**, 1847–1856 (2022).

17. Shen, H. *et al.* Single Particle Tracking: From Theory to Biophysical Applications. *Chem. Rev.* **117**, 7331–7376 (2017).

18. *Handbook Of Biological Confocal Microscopy*. (Springer US, 2006). doi:10.1007/978-0-387-45524-2.

19. Liu, C.-S. *et al.* Novel fast laser-based auto-focusing microscope. in *2010 IEEE SENSORS* 481–485 (2010). doi:10.1109/ICSENS.2010.5690153.

20. Bathe-Peters, M., Annibale, P. & Lohse, M. J. All-optical microscope autofocus based on an electrically tunable lens and a totally internally reflected IR laser. *Opt. Express* **26**, 2359–2368 (2018).

21. Liao, J. *et al.* Single-frame rapid autofocusing for brightfield and fluorescence whole slide imaging. *Biomed. Opt. Express* **7**, 4763–4768 (2016).

22. Zhang, B., Zerubia, J. & Olivo-Marin, J.-C. Gaussian approximations of fluorescence microscope point-spread function models. *Appl. Opt.* **46**, 1819–1829 (2007).

23. Sahoo, S., Davydov, V. A., Agafonov, V. N. & Bogdanov, S. I. Hybrid quantum nanophotonic devices with color centers in nanodiamonds [Invited]. *Opt. Mater. Express* **13**, 191–217 (2023).

24. Quach, N. *et al.* Fluorescence Microscopy Methodology for Visualizing Microscale Interfacial Defects In Packaging Materials. in *2021 20th IEEE Intersociety Conference on Thermal and Thermomechanical Phenomena in Electronic Systems (iTherm)* 1154–1161 (2021). doi:10.1109/ITherm51669.2021.9503140.

25. von Diezmann, L., Shechtman, Y. & Moerner, W. E. Three-Dimensional Localization of Single Molecules for Super-Resolution Imaging and Single-Particle Tracking. *Chem. Rev.* **117**, 7244–7275 (2017).

26. Cang, H., Wong, C. M., Xu, C. S., Rizvi, A. H. & Yang, H. Confocal three dimensional tracking of a single nanoparticle with concurrent spectroscopic readouts. *Appl. Phys. Lett.* **88**, 223901 (2006).

27. Fazekas, F. J., Shaw, T. R., Kim, S., Bogucki, R. A. & Veatch, S. L. A mean shift algorithm for drift correction in localization microscopy. *Biophys. Rep.* **1**, 100008 (2021).

28. Yeo, W.-H. *et al.* Investigating Uncertainties in Single-Molecule Localization Microscopy Using Experimentally Informed Monte Carlo Simulation. *Nano Lett.* **23**, 7253–7259 (2023).

29. Stehr, F. *et al.* Tracking single particles for hours via continuous DNA-mediated fluorophore exchange. *Nat. Commun.* **12**, 4432 (2021).

30. Sutula, M. *et al.* Large-scale optical characterization of solid-state quantum emitters. *Nat. Mater.* **22**, 1338–1344 (2023).

31. Wells, N. P. *et al.* Time-Resolved Three-Dimensional Molecular Tracking in Live Cells. *Nano Lett.* **10**, 4732–4737 (2010).

32. Hou, S. & Welsher, K. A Protocol for Real-time 3D Single Particle Tracking. *J. Vis. Exp. JoVE* 56711 (2018) doi:10.3791/56711.

33. Remmel, M., Scheiderer, L., Butkevich, A. N., Bossi, M. L. & Hell, S. W. Accelerated MINFLUX Nanoscopy, through Spontaneously Fast-Blinking Fluorophores. *Small* **19**, 2206026 (2023).

34. Tan, X. *et al.* Active-Feedback 3D Single-Molecule Tracking Using a Fast-Responding Galvo Scanning Mirror. *J. Phys. Chem. A* **127**, 6320–6328 (2023).

35. Chakraborty, T. *et al.* Converting lateral scanning into axial focusing to speed up three-dimensional microscopy. *Light Sci. Appl.* **9**, 165 (2020).

36. Schneider, J. *et al.* Ultrafast, temporally stochastic STED nanoscopy of millisecond dynamics. *Nat. Methods* **12**, 827–830 (2015).

37. Grimmer, B. Provably Faster Gradient Descent via Long Steps. Preprint at https://doi.org/10.48550/arXiv.2307.06324 (2023).

**Acknowledgements:**

This material is based upon work supported by the U.S. Department of Energy, Office of Science, Office of Biological and Environmental Research, under Award Number DE-SC0023167, and the National Science Foundation NSF-EAGER Grant AK569.

# Supplementary Information

## FiND: Few-shot three-dimensional image-free confocal focusing on point-like emitters

Swetapadma Sahoo[1,2,3], Junyue Jiang[4], Jaden Li[1,5], Kieran Loehr[3,5], Chad E. Germany[3,5], Jincheng Zhou[4], Bryan K. Clark[3,5], Simeon I. Bogdanov[1,2,3]

[1]Department of Electrical and Computer Engineering, University of Illinois at Urbana-Champaign, Urbana, Illinois 60801, USA

[2]Nick Holonyak, Jr. Micro and Nanotechnology Laboratory, University of Illinois at Urbana-Champaign, Urbana, Illinois 61801, USA

[3]Illinois Quantum Information Science and Technology Center, University of Illinois Urbana-Champaign, Urbana, Illinois 61801, USA

[4]Zhejiang University-University of Illinois at Urbana-Champaign Institute, Haining 314400, China

[5]Department of Physics, University of Illinois Urbana-Champaign, Urbana, IL 61801, USA

**Supplementary Note 1: Analytical approximation of the resultant force**

The total displacement **D** is the vector sum of the ground truth component **D$_{GT}$** and the noise component **D$_N$**. The ground truth displacement is given as:

$$\mathbf{D_{GT}} = -\frac{\lambda}{\delta}\sum_{j=x,y,z} e^{-\frac{j^2}{2}-\frac{\delta^2}{2}} \sinh(j\delta)\, \mathbf{e}_j \quad (S1)$$

**D$_{GT}$** is directed essentially radially, but a transverse component might be present as **D$_{GT}$** is generally not colinear with **r**. However, the comparison of analytical resultant force in Fig. 2 of the main text with the Monte Carlo simulations shows that neglecting the transverse **D$_{GT}$** component is a good approximation.

Next, we notice that the variance of **D$_N$** projection on any direction **e** does not depend on **e**:

$$\langle (\mathbf{D_N} \cdot \mathbf{e})^2 \rangle = \left\langle \left(\frac{\lambda}{2\delta}\sum_{j=x,y,z}\{n(\mathbf{r}+\delta*\mathbf{e_j},t) - n(\mathbf{r}-\delta*\mathbf{e_j},t)\}\mathbf{e}_j \cdot \mathbf{e}\right)^2 \right\rangle = \frac{\lambda^2 \bar{n}^2}{2\delta^2} \quad (S2)$$

where **e** is an arbitrary vector of length 1. $n(\mathbf{r}+\delta*\mathbf{e_j},t)$, $n(\mathbf{r}-\delta*\mathbf{e_j},t)$ are normally distributed independent variables, so their difference has a variance of $2\bar{n}^2$

By taking the projection of **D$_N$** onto the radial direction, we get $\langle |\mathbf{D_{RN}}|^2 \rangle = \frac{\lambda^2 \bar{n}^2}{2\delta^2}$. Similarly, the variance of the transverse noise component is determined by the sum of the variances of its two projections on axes orthogonal to **r**, which are both equal to $\frac{\lambda^2 \bar{n}^2}{2\delta^2}$. Therefore, $\langle |\mathbf{D_{TN}}|^2 \rangle = \frac{\lambda^2 \bar{n}^2}{\delta^2}$.

We now consider the focus of the beam as a particle moving along the effective radial coordinate r$^2$ (Fig. 2 of main text). The forces are defined as the average increment in the coordinate r$^2$ as a result of one iteration. Using $\langle \mathbf{D_N} \rangle = 0$, we express the resultant force in terms of radial and transverse displacements as:

$$F_{res}(r) = \langle |\mathbf{r}+\mathbf{D}|^2 - r^2 \rangle = D_{GT}^2 + 2rD_{GT} + \langle |\mathbf{D_{RN}}|^2 \rangle + \langle |\mathbf{D_{TN}}|^2 \rangle \quad (S3)$$

By using the expressions of the displacement variances, we arrive at the equation (1) in the main text:

$$F_{res}(r) = -2r\frac{\lambda}{\delta}e^{-\frac{r^2}{2}-\frac{\delta^2}{2}}\sinh(r\delta) + \frac{\lambda^2}{\delta^2}e^{-r^2-\delta^2}\sinh^2(r\delta) + \frac{3\lambda^2\bar{n}^2}{2\delta^2} \quad (S4)$$

**Supplementary Note 2: Finding optimal values of $\delta, \lambda$ that maximize the theoretical resultant force**

To find the optimal $\delta$ at a given r, it is convenient to parameterize the expression of the resultant force (Eq. S4) using $\lambda = K\delta$ (where K is a constant):

$$-F_{res}(r) = 2rKe^{-\frac{r^2}{2}-\frac{\delta^2}{2}}\sinh(r\delta) - K^2 e^{-r^2-\delta^2}\sinh^2(r\delta) - \frac{3K^2\bar{n}^2}{2} \quad (S5)$$

Maximizing -$F_{res}$(r) gives $\delta \approx 1$ when r << 1, and $\delta = r$ when r >> 1.

To rapidly converge towards focus, FiND benefits from sampling high-ground truth information regions of the noisy Gaussian function. Near the intensity peak (r << 1), this corresponds to an optimal step size of approximately 1σ. Conversely, when distant from focus (r >> 1), a step size on the order of rσ is required to sample intensities in this region.

To calculate the parameters yielding the fastest focusing time, we maximize the resultant force at the target zone boundary, -$F_{res}$ ($r_\epsilon$) (Fig. 1 of main text). For $r_\epsilon$ =0.44, we choose $\delta$= 1.

Maximizing -$F_{res}$(r) (Eq. S4) along $\lambda$, with the assumption of $\delta$= 1 and small r's:

$$\lambda \sim \sqrt{e}\frac{2r^2}{2r^2+3\bar{n}^2 e} \quad (S6)$$

**Supplementary Note 3: Comparison of FiND in finite-difference sampling framework against several optimization algorithms**

We compare the performance of FiND in finite difference sampling framework to that of several optimization algorithms whose parameters were optimized using an iterative grid search which became more finely tuned after each step. Specifically, we consider the natural evolution strategy (NES), particle swarm optimization (PSO) optimization for iterative focusing, and convolutional neural network (CNN)-based curve fitting for non-iterative focusing. We compare noise-robustness (Fig. S1(a)) and focusing speeds (Fig. S1(b)) through Monte-Carlo simulations.

NES is a second-order optimization method that tracks the function's natural gradient[1]. It generates a batch of search points, estimates the natural gradients, and performs small steps along these gradients. It excels when optimizing functions with rapidly changing curvatures because it accounts for the second derivative, in addition to the first derivative. NES was implemented using standard methodology, with the 'default' 3-dimensional gaussian search distribution with variable mean and the same fixed standard deviation (of 1, same as FiND simulations) in 3 dimensions[1]. We optimized three key parameters: the standard deviation of the search distribution, the learning rate, and the population size via iterative grid search. However, it's important to note that the time-varying noise will also affect the estimation of the curvature and thus amplify the impact of time-varying noise compared to a first-order method like FiND.

PSO uses a population of candidate solutions (particles) and updates the particle positions in the search space[2]. Each particle moves according to its own best-known positions in space as well as the globally best-known position, ultimately converging to the function's optimum. The parameters optimized via iterative grid search were the number of particles, gradient-based learning rate, and acceleration coefficients. In cases of low SNR PSFs, noisy data may mislead the algorithm by prematurely identifying suboptimal points, preventing exploration of the true optimal region.

An AlexNet-like CNN regression model used a multilayered 1D CNN with two convolutional layers, two max-pooling layers, followed by three dense layers and a flatten layer to curve fit and predict the position of maximum intensity from a batch of sampled intensities in the 3D space[3]. The network was trained on three million simulated noisy gaussian functions with known coordinates of highest intensity. The intensities were sampled using a uniform grid for each gaussian function. We verified the performance of the trained CNN by analyzing its predicted coordinates of highest intensity after sampling previously unseen noisy gaussian functions. Parameters optimized were learning rate and activation function. While such regression models perform well for fitting noiseless functions rapidly with sparsely sampled data[4], we observe that it struggles to fit under high-noise conditions[5]. It remains to be seen if modifying the CNN structure by adding more layers, regularizing the model, and further optimizing the hyperparameters can improve the results.

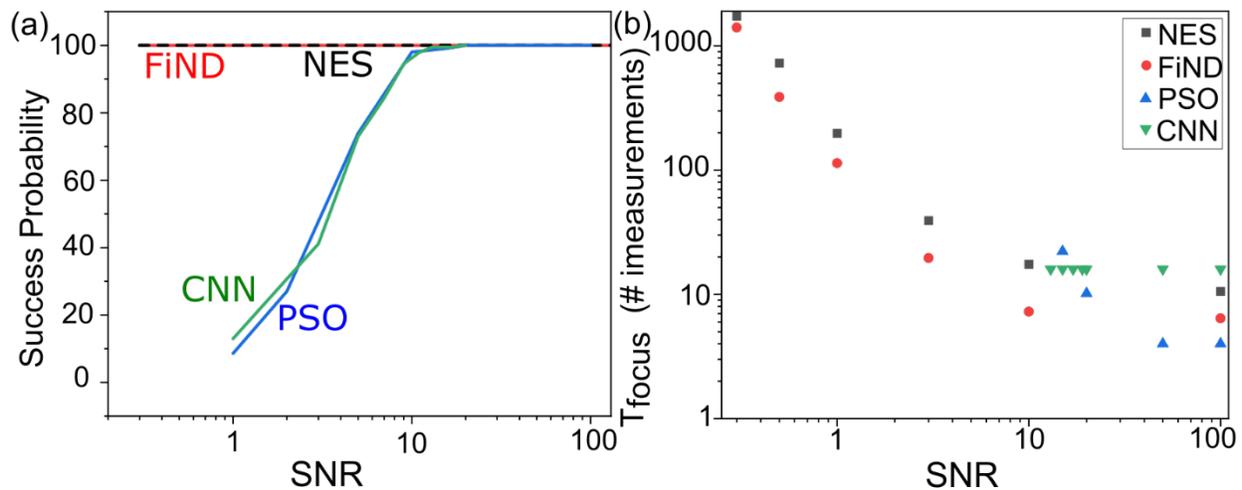

**Fig. S1.** Numerical results comparing the performance of FiND with focusing algorithms based on NES, CNN and PSO. (a) The success probability of focusing is evaluated for SNRs>0.3. It was found that FiND (red curve) and NES (dashed black curve) could successfully focus for all the SNRs. CNN (green curve) for SNRs>19 and PSO (blue curve) for SNRs> 100. (b) For all the successful focusing attempts in (a), each algorithm's total number of measurements is plotted. FiND performs the best, having the least number of measurements required for successful focusing at SNRs>0.3.

**Supplementary Note 4: FiND focusing on blinking emitters**

Since our focusing algorithm has no memory, it can focus on emitters featuring a pronounced blinking behavior. Fig.S2 shows two such example instances. We observe that FiND attains focus by entering the threshold of intensities > 0.9.

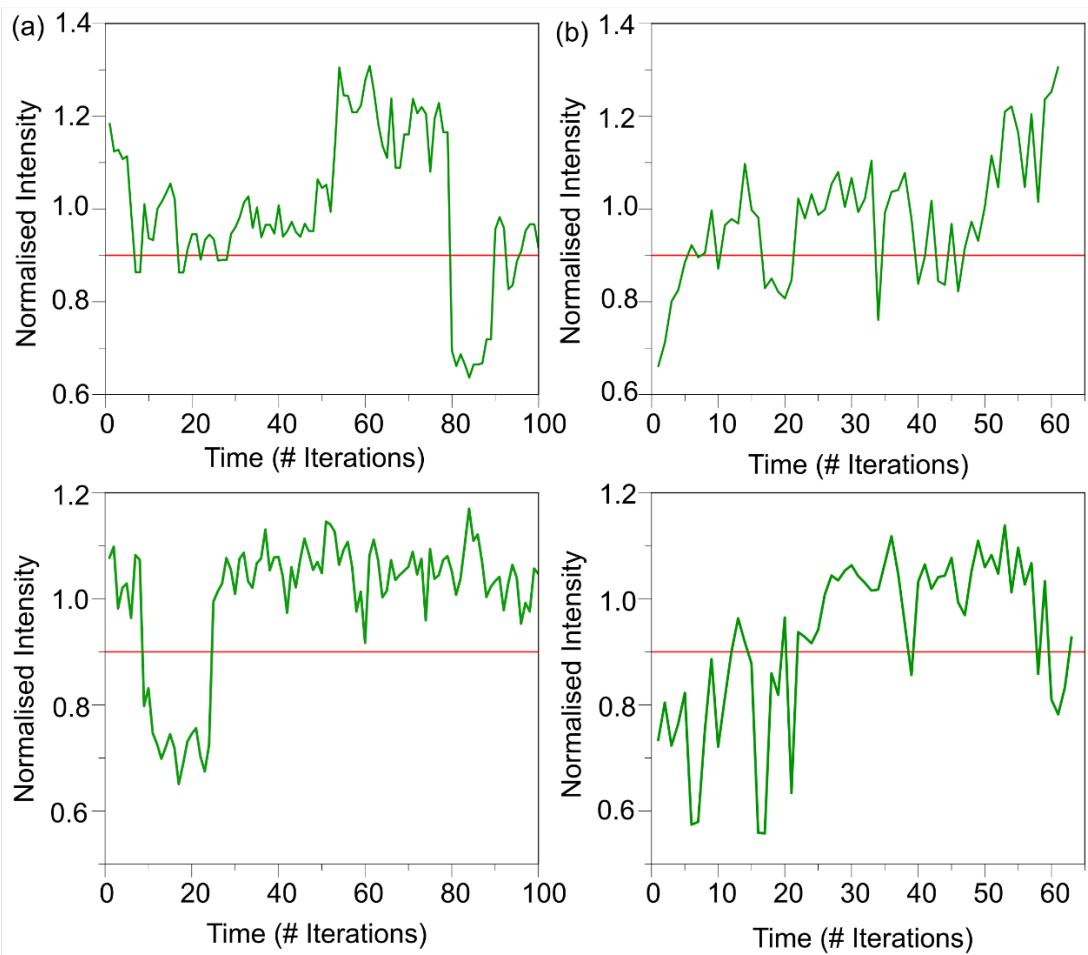

**Fig. S2.** FiND focuses on blinking emitters (a), (c) Stability curves of the emitter show photoblinking. (b), (d) FiND focusing curve shows that focusing is attained entering the threshold of intensities > 0.9 (red horizontal line).

**References:**

S1. Wierstra, D., Schaul, T., Peters, J. & Schmidhuber, J. Natural Evolution Strategies. in *2008 IEEE Congress on Evolutionary Computation (IEEE World Congress on Computational Intelligence)* 3381–3387 (IEEE, 2008). doi:10.1109/CEC.2008.4631255.

S2. Miranda, V. & Fonseca, N. EPSO-evolutionary particle swarm optimization, a new algorithm with applications in power systems. in *IEEE/PES Transmission and Distribution Conference and Exhibition* vol. 2 745–750 vol.2 (2002).

S3. Singh, I., Goyal, G. & Chandel, A. AlexNet architecture based convolutional neural network for toxic comments classification. *J. King Saud Univ. - Comput. Inf. Sci.* **34**, 7547–7558 (2022).

S4. Austin, P. C. & Steyerberg, E. W. The number of subjects per variable required in linear regression analyses. *J. Clin. Epidemiol.* **68**, 627–636 (2015).

S5. On Fitting a Straight Line to Data when the "Noise" in Both Variables Is Unknown in: Journal of Atmospheric and Oceanic Technology Volume 30 Issue 1 (2013). https://journals.ametsoc.org/view/journals/atot/30/1/jtech-d-12-00067_1.xml.